# The German eID as an Authentication Token on Android Devices

Florian Otterbein[#1], Tim Ohlendorf[*2], Marian Margraf[#3]

[#]*Department of Mathematics and Computer Science, Freie Universität Berlin*
*Takustrasse 9, 14195 Berlin, Germany*
[1] `florian.otterbein@fu-berlin.de`
[3] `marian.margraf@fu-berlin.de`

[*]*Department of Computer Science, Technische Universität Darmstadt*
*Hochschulstrasse 10, 64283 Darmstadt, Germany*
[2] `tim.ohlendorf@stud.tu-darmstadt.de`

*Abstract*—Due to the rapid increase of digitization within our society, digital identities gain more and more importance. Provided by the German eID solution, every citizen has the ability to identify himself against various governmental and private organizations with the help of his personal electronic ID card and a corresponding card reader. While there are several solutions available for desktop use of the eID infrastructure, mobile approaches have to be payed more attention. In this paper we present a new approach for using the German eID concept on an Android device without the need of the actual identity card and card reader. A security evaluation of our approach reveals that two non-critical vulnerabilities on the architecture can't be avoided. Nevertheless, no sensitive information are compromised. A proof of concept shows that an actual implementation faces some technical issues which have to be solved in the future.

## I. INTRODUCTION

Germany introduced the electronic ID card (eID) on November, 1st 2010. In combination with a desktop client and an RFID card reader, it enables citizens and legal residents to identify themselves online against private and governmental organizations (eID offerer). However, a consumer research from 2015 determined that only about 5% of all Germans used their eID for online authentication services within the past 12 months [12].

We can think of two important reasons. First, only few services with eID support are available on the market. Thus, users may not see a significant benefit in using the eID. Second, ID card holders feel inconvenient in using the eID. A user study by Willomitzer et al. figured out that using the required card reader and installing the necessary drivers are one of the biggest usability problems and result in a high cancellation rate [29].

In this paper, we present an eID concept that eliminates the requirement of a card reader by relying on the user's smartphone instead of an ID card. During an initial setup phase, an eID token will be installed on the users smartphone. Online authentication services can then be used without the user's ID card. This could increase the acceptance rate of eID among citizens and legal residents.

This paper is structured as follows. First, we present related work and an outline in Sec. II. In Sec. III we will introduce the technologies that are used for our architecture in Sec. IV. In Sec. V, the results of our security evaluation are presented. Finally, we will discuss our prototype implementation in Sec. VI and summarize all results in Sec. VII.

## II. RELATED WORK

Existing projects provide a mobile application for the German eID by using the internal NFC controller, an external Bluetooth card reader, or an USB card reader [21, 22]. Besides the lack of usability by using an external device during the authentication process, Android's NFC stack is unable to handle extended Application Protocol Data Unit (APDU) commands [18]. As the eID process uses extended APDUs for communication, it is incompatible with Android smartphones. Additionally, existing eID applications only initiate the authentication process and act as a management component. All security relevant operations of the process are performed by the smart card chip inside the ID card.

Schröder developed and evaluated a concept that enables storing and using derived identities on a smartphone securely [26]. He demonstrated his approach in a proof of concept with the German eID and a corresponding wireless card reader. While this is important work and describes details about managing identities on smartphones, it still relies on a card reader. In Sec. I we mentioned, that the card reader is one of the biggest usability problems. Therefore the proposed solution could be improved.

Other countries in the European Union also provide mobile authentication solutions for their citizens. In Austria for example one can register for the so called "Handy-Signatur" [1] which is a mobile signature system offered by the Graz University of Technology, the Austrian Government and Austrian National Bank. This project, initialized 2009, first used SMS-TANs for authentication. Today a corresponding smartphone app can be used instead. Another solution for mobile authentication by the Austrian state printing office is "My Identity

App" (MIA) [24]. MIA stores all governmental identification documents, like passport, drivers license, etc. centralized in a digital format.

Both Austrian approaches have in common, that they require a cloud backed infrastructure. All sensitive personal information is processed and stored on the provider's servers. This stands in contrast to our following approach were the sensitive data is never revealed to a third-party.

Apple Pay, a mobile payment service, serves as an example for a successful usage of digital identities. A survey from 2015 showed a high acceptance rate, with 39% of all iPhone 6 users in the US and UK already using it and an additional 30% of participants who would like to use it in the future [17].

### III. TECHNICAL BACKGROUND

Saving a sovereign identity on a mobile device leads to new security requirements. The storage and transmission of an identity to the eID offerer has to be protected with a similar security level as on eID cards which use a secure element.

Thus we rely on the secure element and trusted execution environment of an Android phone that are described in this section. Also a short overview of the underlying eID process is given.

#### A. Online Authentication (eID process)

The German eID system enables the owner of a German ID card (eID) to identify himself against various service providers (eID offerer) in the governmental and private sector. Therefore all components involved in the eID process have to fulfill certain requirements described in technical guidelines, published by the German Federal Office for Information Security (BSI) [5]. The security and privacy aspects of the used protocols are addressed in various papers and articles [3, 4].

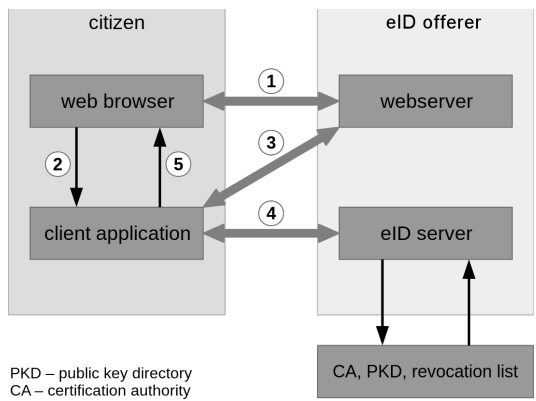

1. eID process between citizen and eID offerer

Fig. 1 shows a schematic representation of an eID process which will be explained in the following:

- The user wants to identify himself against an eID offerer. Therefore he clicks on an special hyperlink (TcTokenURL) on the website of the eID offerer.
- The user's browser forwards the URL to the eID client application.
- The client software recognizes the TcTokenURL and fetches the request (TcToken) containing the required personal data of the user (name, surname, date of birth, ...) and the responsible eID server for the current eID process.
- The eID client establishes a secure connection to the eID server and starts the authentication process between the eID card and the eID server. Finally the id card sends the required personal data to the eID server.
- The eID offerer receives the needed personal data of the user from the eID server. The user is redirected to an authenticated web session with the eID offerer.

To ensure a secure connection between eID card and eID server the EAC (Extended Access Control) protocol [5] is used. It authenticates the two involved participants against each other and is divided in two subprotocols called Terminal Authentication (TA) and Chip Authentication (CA). TA guarantees a valid and authentic communication between eID card and eID server. CA guarantees the eID server to communicate with a valid and authentic eID card. To provide this, Public-Key-Infrastructures are used (Country Verifying Certificate Authority (CVCA) and Country Signing Certificate Authority (CSCA)) with CVCA and CSCA certificates as root instances. The underlying cryptographic primitive is a Diffie–Hellman key exchange, the public keys of both parties are authenticated using signature algorithms.

#### B. Secure Element

A Secure Element is a tamper-resistant microcontroller that provides a secure environment for applications and data [16]. Its computing power and storage capacity are limited and can be found on the Universal Integrated Circuit Card (UICC), SD card or on a non-replaceable chip embedded in the smartphone (see Fig. 2).

The secure element is divided into security domains that can be programmed using Java Card applets [23].

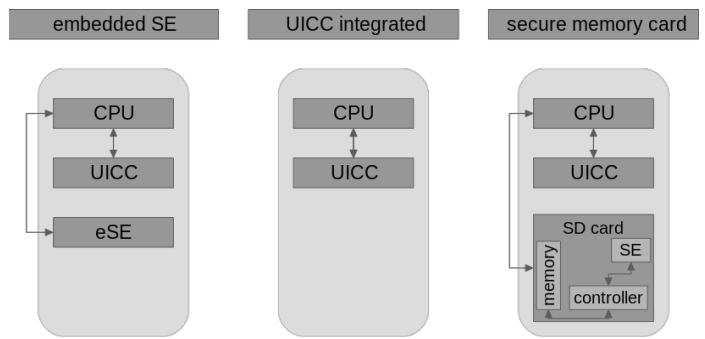

2. Secure Element form factor overview on mobile devices

A Trusted Service Manager (TSM) is the link between the secure element manufacturer (issuer) and the application developer (service provider). The TSM is responsible for provi-sioning and personalizing the secure element. Without a TSM, the service provider must have a contract with every issuer.

*C. Trusted Execution Environment*

A Trusted Execution Environment (TEE) enables the execution of security-critical applications inside a runtime environment separated from the rest of the system [28]. This isolated environment can be implemented as an external coprocessor or as different runtime modes on the main CPU. Many microprocessors on the today's market (e.g. ARM, AMD, Intel) feature a TEE.

Focusing on mobile systems, ARM CPUs are currently most relevant for this consideration. In ARMs TEE implementation TrustZone [2], trustlets are security-critical applications that can be executed in an isolated environment (secure world), while the actual Rich OS is running in a non-isolated environment (normal world), see Fig. 3. Both worlds have their own virtual allocated storage area on the physical RAM and in the registers of the CPU. Communication between the worlds is managed by a monitor, located in the secure world and waiting for requests (Secure Monitor Calls) from the normal world.

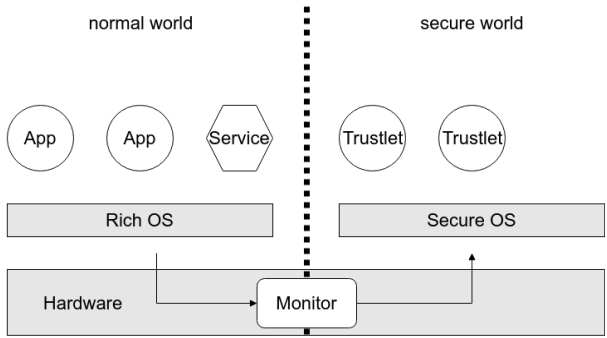

3. Schematic representation of the ARM TrustZone architecture

IV. DESIGN

*A. Initialization*

In this section the architecture and installation process is described. Issuer, TSM and Service Provider each have a local instance (security domain/applet) and a remote instance (external server). During the smartphone's manufacturing process, only the security domain of the issuer is installed. The initialization procedure is based on GlobalPlatform specifications [14]. Figure 4 shows all instances and their connections.

1) *Initial setup:* The user installs the eID app from Google's Play Store. The app checks if a TEE and a secure element are available and if so, it installs the trustlet for the TEE.
2) *Installing TSM's security domain:* The Android app sends a request for installing a new security domain (also called supplementary security domains (SSD)) to TSM's remote instance. The TSM forwards the command to the remote instance of the issuer.
   Local and remote instance of the issuer authenticate each other by using a challenge-response protocol.

The issuer's remote instance sends an encrypted installation command to his local instance and sets new access keys for the TSM. The issuer sends the new keys to the TSM, who can directly connect to his new local instance for setting his own keys.

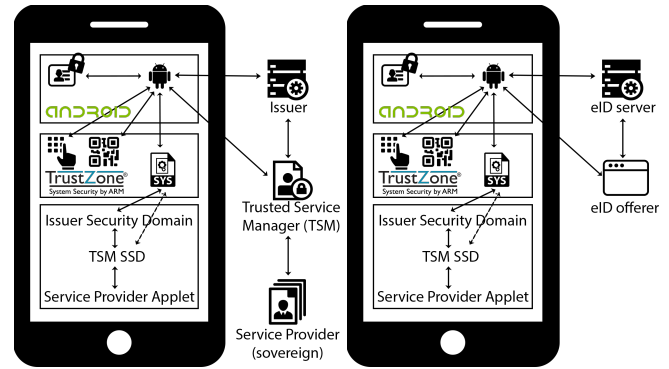

4. Mobile eID architecture with Service Provider, TSM, Issuer, eID server, eID offerer, TEE, and secure element.
Left side: Actors and communication channels during initialization process.
Right side: Actors and communication channels during authentication process.
Inside the TEE (e.g. ARMs TrustZone), users can enter secret information. The TEE also works as a bridge between the non-isolated environment and the secure element. Inside the secure element, isolated security domains and applets are implemented that are managed by their external instances. Dashed lines show that this communication channel can only be used for setting new encryption and authentication keys, based on GlobalPlatform's definition [14].

3) *Installing the applet:* The implementation of protocols and installation of certificates is needed to support the eID process. Thus an applet for the Android app and the root certificates CVCA and CSCA have to be installed. The TSM's remote instance sends an encrypted and signed applet to the issuer, whose local instance is responsible for installing the applet. TSM's local instance has to verify the applet to release the installation process. This verification process is called DAP-Verification [14]. Afterwards the service provider has its own non-personalized applet inside the security domain.

4) *Personalize security domain:* To protect the Android app against abuse, an access PIN is needed, which is define by the user during the personalization process.
   A key exchange is used to establish a secure connection between the service provider and its applet. As Android's NFC stack is incompatible with German eID (see Sec. II), we use the smartphone camera for capturing the ID card. This process is legally compliant with the German Money Laundering Act [7] and is already practiced by some German-based companies.
   Beside the traditional PIN of the ID card, a personal private key (represented by a QR-Code) has to be added to the postal letter of the sovereign service provider, that is sent to the citizen after ordering an ID card.
   By capturing the ID card and this key, the mobile eID has the same security level as the traditional eID process (see Sec. V). The external service provider receives the captured data and checks its integrity. If the validation process is successful and the ID card wasn't used for a smartphone initialization process before (this avoids cloning eID tokens), the service provider sends back the eID token containing all personalized information, concatenated with the private key of the Chip Authentication (see Sec. III-A) to the security domain. The data package is encrypted with the public key of the QR-Code. The applet can decrypt the token, encrypt it with a symmetric key and store the data on Android's file system. This is necessary as the smart card only provides little storage capacity. The private key for the Chip Authentication remains inside the secure element.

5) *Authentication process:* After the installation process, the eID token is installed and the ID card is not needed anymore. The eID token's decryption key can only be accessed by entering the PIN into the TEE. After retrieving the key, the secure element in the smartphone can act like the secure element in the ID card.

### B. Authentication

The authentication protocols (see Sec. III-A) used by the eID app are determined in TR-03110 [5]. Some modifications regarding the access policy are necessary. The communication channels and all actors are shown in Fig. 4.

- The user clicks on the eID offerer's link. A secure connection between Android app and eID server is established.
- By entering the PIN into the TEE, the access for the user to the secure element is granted.
- The terminal authentication between secure element and eID server is carried out.
- The chip authentication between secure element and eID server is carried out.
- The encrypted token in the untrusted system storage is received by the secure element and gets decrypted by the key stored inside the secure element.
- The user determines what kind of personal data inside the TEE should be redirected to the eID server.
- Secure element and eID server can directly communicate using encrypted and authenticated APDU commands.
- Once the authentication process is finished, the access to the secure element is locked again.

## V. SECURITY EVALUATION

The evaluation is based on the design in Sec. IV and will be divided into the following steps:

1. Initialization: Description of business processes and security objectives
2. Structural Analysis: Description of actors, components and communication channels
3. Security Requirements: Definition of assets that need to be protected and analysis of their protective needs
4. Modeling of IT Architecture: Summary of actors and components
5. Description of Protective Measures: Protective measures are established to ensure the security of assets that need to be protected. Specific security functions are used to implement the protective measures.
6. Evaluation of Protective Measures

### A. Initialization

We consider the use cases *initialization* and *authentication*. A description of these use cases can be found in Sec. IV-A and IV-B, respectively.

We will distinguish between the security objectives confidentiality, authenticity and availability:

1. Confidentiality: Protecting data from disclosure to unauthorized parties
2. Authenticity: Ensure, that data was not sent by unauthorized parties (includes integrity).
3. Availability: Ensure, that data can be accessed by authorized parties when needed.

### B. Structural Analysis

All actors and communication channels are described in Sec. IV. We assume, that the existing protocols for the authentication process are secure. Security evaluations regarding the communication between smartphone, eID offerer and eID server already exist [25]. We do not need to evaluate the communications between two external instances for the same reason.

Therefore, it is sufficient to investigate the following components, that are necessary to get the personalized token on the smartphone at the initial setup (use case initialization) and using the personalized token for authentication (use case authentication):

- Smartphone including all components and their communication
  - Android app
  - Internal storage of Android app
  - Host CPU
  - TEE
  - Secure element
- Issuer
- TSM
- Service Provider
- Communication between Android app and security domains (over TEE)
- Communication between Android app and issuer
- Communication between Android app and TSM

### C. Security Requirements

Due to the high protection needs of the business process "authentication" (see Sec. IV), the asset "eID token" has high protection needs and thereby the smartphone and the components being in contact with this asset as well. Furthermore, we can determine the protection needs by analyzing the asset "eID token" regarding their specific security objectives:

- Confidentiality: High protection needs regarding confidentiality, because the eID token contains personalized data. By having access to all data stored inside the token, an attacker could harm the user.
- Authenticity: High protection needs regarding authenticity, because the business process aims to authenticate the user with the help of the eID token on different services. For examples, using the eID token for opening a bank account requires a high security level.
- Availability: Normal protection needs regarding availability. If an authorized user isn't able to access the eID token when it's needed, he can't authenticate online and loses comfort.

It can be concluded, that an attacker can cause high damage by manipulating or having access to another eID token. In combination with the high spread of Android malware [10] and thereby a high probability of being attacked, a high risk exists.

### D. Modeling of IT Architecture

As a summary of Sec. IV, Fig. 4 shows all actors and components during the initialization and authentication process.

### E. Description of Protective Measures

In this section we will describe the protective measures that are necessary to fulfill the security requirements mentioned in Sec. V-C.

1) *Keys of issuer:* Issuer keys are needed to ensure a confidential and authentic communication between issuer and its security domain. During the manufacturing process, the symmetric keys S-ENC and S-MAC are stored inside the issuer's security domain. The keys don't leave the secure element. The issuer and its security domain calculate a session encryption key by encrypting a random number with the key S-ENC. To ensure the authentication of messages sent between issuer and its security domain, the cryptographic algorithm HMAC with the key S-MAC is used.
2) *Keys of TSM:* After the issuer installs a security domain for the TSM, the access keys of the security domain are transferred from the issuer to the TSM. The TSM can now access his new security domain over an confidential and authentic communication channel. The GlobalPlatform specification [14] allows a connection without the help of the issuer (see dotted lines in Fig. 4) if the TSM sets new keys. Thereby the issuer can't manipulate or tap further communication between TSM and its security domain.
3) *Access PIN for Android application:* Only the smartphone owner should use his credentials stored inside the smartphone to authenticate against an eID offerer. The user enters his PIN inside the TEE to ensure, that no attacker can tap his inputs. The PIN is evaluated inside the secure element by the internal Java Card applet of the service provider.
4) *Personal private key (QR code):* Nobody should be allowed to store the eID token of another person on his smartphone. A two-factor authentication is used during the initialization process to ensure only the ID owner is able to use the eID token. An additional QR code, which represents a private key, is printed on the letter received by the ID owner. The associated public key is used by the service provider to encrypt the eID token. The user scans his QR code inside the TEE. To decrypt the received eID token, the personal private key is necessary.
5) *eID token's symmetric encryption/decryption key:* The limited resources of the secure element prevent the persistent storage of the eID token. Therefore the token has to be stored on Android's file system. A symmetric key is generated and stored on the secure element to encrypt and decrypt the eID token.

### F. Evaluation of Protective Measures

The evaluation of the used cryptographic algorithms and physical attacks against the secure element are excluded. It's assumed that the TSM is trustworthy, the attacker has limited resources and no implementation errors have been made.

All commands that are executed on the Host CPU of the Android smartphone are particularly critical, because of high attack potential. As a result, two vulnerabilities have been found:

1) *Sniffing SELECT commands:* The SELECT command is responsible for addressing the correct security domain inside the secure element. The GlobalPlatform organization is responsible for defining secure element standards. In their definition about the communication, it is said, that no encryption of the SELECT command is possible. Thus it is possible to obtain information about the identification number of the security domain. No harmful attacks can be executed with this information, because the security domain can only be accessed by the eID Android application and with an additional PIN.
2) *Relay attack:* During the installation of the security domain from the TSM or service provider, an attacker could redirect the commands between Android app and the secure element. If a user compares the installation process of the security element in a TEE with the installation process in a rich execution environment, it would be possible to detect a successful attack in the past. No sensitive information is compromised with this attack. During the personalization process, the transferred token can also be redirected. Due to cryptographic mechanisms, attackers can't decrypt the sensitive information.

## VI. IMPLEMENTATION

As the implementation is still work in progress, an overview of the successfully finished tasks will be given. The fol-lowing is divided into two subsections, addressing different aspects of the provided architecture in Sec. IV. With the help of the Android prototype in Sec. VI-A, usability aspects regarding the eID process on Android smartphones can be analyzed. Sec. VI-B lists requirements smartphones must meet to support our mobile eID solution.

### A. Usability Prototype

The native Android prototype was developed for usability testing and does not provide any application logic for dealing with a real identity card. It emulates the specified eID functionality (initialization, eID process), shown in Fig. 5, by serving dummy data which gives the possibility to monitor user behavior in lab tests. Additionally a feature to find supported eID offerer in the Google Play Store or on the web was added to the prototype as well.

The usability of our implemented application has been evaluated by Kostic et al. [19].

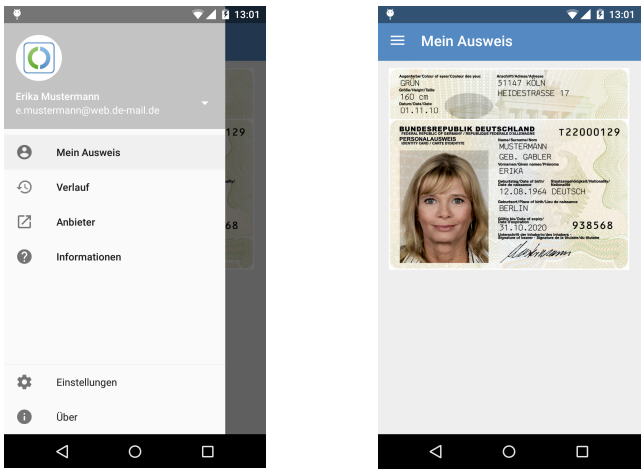

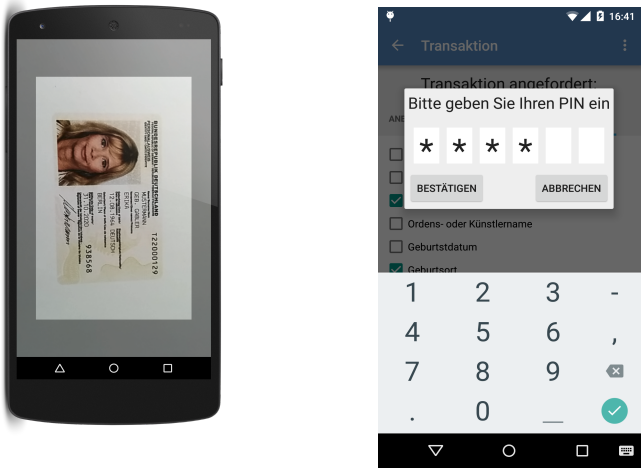

5. Usability prototype of the eID application on Android.
Upper left: Main menu, upper right: my ID card
Lower left: initialization, lower right: eID process

### B. Technical Evaluation

The architecture in Sec. IV makes use of two hardware components (ARM TrustZone and secure element) to strengthen the security of the mobile eID concept on Android. In the following, existing technical solutions for Android are evaluated. First, the requirements which have to be fulfilled by each component will be mentioned. Then, existing solutions will be provided. Finally, a proof of concept will be discussed.

1) *ARM TrustZone*

   *a) Requirements:* Access to the ARM TrustZone on Android devices as well as the ability to install, manage and remove developed trustlets.

   *b) Existing Solutions:* As described in Sec. III-C, most current Android smartphones on the market provide an ARM TrustZone. The bare Android OS itself only uses the TrustZone for key generation / storing and DRM purposes. Some smartphone manufacturer like Samsung also integrate real-time kernel protection and secure application environments (e.g. KNOX) into their custom Android and TrustZone versions [20]. The communication between the Android OS and the trustlets inside the TrustZone is commonly managed by a binary blob driver in the kernel provided by each smartphone/chip manufacturer. At the time of writing, third-party developers who want to make use of the TrustZone are locked out. There are currently no official mechanisms provided for third-party developers to install their own trustlet. However, the GlobalPlatform association is currently working on standards for implementing and managing TEEs and enabling third-party developers to access the secure world for use in their application [15].

   *c) Proof of concept:* Because of missing access to the TrustZone, a proof-of-concept was not developed. However the implementation of trustlets in general was addressed in various papers in literature [27, 30].

2) *Secure Element*

   *a) Requirements:* Access to a secure element installed on the Android smartphone. Deploying, installing and managing a security domain for the eID applet. Support of needed cryptographic algorithms for the eID process.

   *b) Existing Solutions:* Secure elements on mobile devices exist in different form factors (see Sec. III-B). Focusing on Android smartphones the most common is the UICC. Google replaced the embedded secure element in their developer phones with a software emulator called Host-based Card Emulation when their Nexus series has been released in 2013 [8]. Therefore, the embedded secure element is no longer officially supported by the Android API [9]. The manufactures of removable SD card secure elements face these issue by developing their own proprietary drivers, based on the seek-for-android open-source project [13]. All available solutions have in common, that they are currently not accessible for third-party developers.

   *c) Proof of concept:* Based on the presented concept, the secure element should perform the EAC protocol (see Sec. III-A) and should store the eID token. For the proof-of-concept we used a SD card secure element (certgate cgCard Type 2) which is accessible via a customized middleware based on seek-for-android and open for developers. Further experiments were not possible because of missing cryptographic algorithms [6] (ALG_EC_SVDP_DH_PLAIN for key agreement). Secure elements with the JavaCard 3 Java platform and a modified JCOP SE OS (v2.4.1) should fix this issue, but were not available at the time of writing.

3) *eID Android Application*

   *a) Requirements:* Application logic to manage eID process and communicate with the hardware security components (TrustZone, secure element).

   *b) Existing Solutions:* Currently, two open source projects (Open eCard [22] and PersoApp [21]) provide a mobile Android application for the German eID system. Open eCard also supports further identity cards like the German and Austrian electronic health insurance card. Both applications communicate via different card readers with the identity cards.

   *c) Proof of concept:* As the PersoApp focuses exclusively on the German eID card, it will be used for further experiments. It uses a modular design [21] to support different kinds of card readers (Bluetooth, USB, NFC) via the so called CardHandler interface. By introducing a new CardHandler in Fig. 6 (SECardHandler) and emulating an identity card connected to a card reader with a PIN pad, we could easily use the PersoApp without any further modifications. Therefore the SECardHandler communicates over the SEService and the eID emulator with the secure element. In this design, the TrustZone is left out because of the restrictions mentioned earlier, but could be integrated into the SEService in the future.

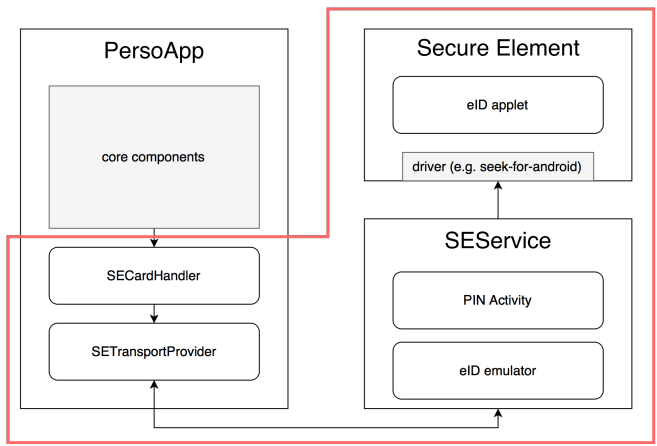

6. Modified PersoApp with SECardHandler, SEService and Secure Element (red box)

C. *Conclusion*

The evaluation of the concept revealed some issues which must be solved before an implementation and a final product could be released.

The lack of a TrustZone API for third-party developers makes it impossible to use it as defined in the concept. Nevertheless, if the GlobalPlatform TEE API gets deployed on Android devices in the future, another evaluation should be done.

As embedded secure elements are no longer supported on Android by default and third-party SD card secure elements need additional custom drivers, a possible method of choice could be UICCs as they are already distributed in every smartphone. The existing management infrastructure of the mobile network carriers could be adapted to deploy the eID applet onto the devices. Also the problem of missing algorithms can be fixed by using UICCs which support JavaCard 3 (e.g. Gemalto UpTeq Multi-Tenant SIM [11]).

Finally, the proof of concept for the Android application shows no evidence for possible incompatibilities with the concept.

IV. CONCLUSION AND FUTURE WORK

In this paper we demonstrated that hardware-based security solutions are needed to protect sensitive data. Smartphone owners with custom flashed kernels and root privileges are able to obtain sensitive information, if only software-based security mechanisms are used. The security evaluation showed, that two non-critical vulnerabilities can't be avoided. Neither the eID token nor the cryptographic keys were compromised.

It's possible to transfer the described concept to eID solutions of other countries as long as a sovereign service provider managing the eID tokens is available and the processes are compliant with the national law.

Nevertheless, it is complicated to enroll the concept to Android because of the heterogeneous device hardware: Each device needs a secure element and a trusted execution environment which is able to understand the APDU commands sent by the service provider, support needed algorithms for the eID process and provide an API for third-party developers.

As mentioned in Section VI-C, future work for the implementation should focus on UICCs as the secure element. Their widespread in almost every smartphone and the existing management infrastructure of the mobile network carriers could be adapted for the mobile eID concept. Also the use of the ARM TrustZone must be reevaluated when manufacturers include the GP TEE API [15] into their implementations.